\documentstyle[12pt]{article}
\advance \textheight by \topskip
\makeatletter
\def\eqnarray{\stepcounter{equation}\let\@currentlabel=\theequation
\global\@eqnswtrue
\global\@eqcnt\z@\tabskip\@centering\let\\=\@eqncr
$$\halign to \displaywidth\bgroup\@eqnsel\hskip\@centering
  $\displaystyle\tabskip\z@{##}$&\global\@eqcnt\@ne 
  \hfil$\displaystyle{\hbox{}##\hbox{}}$\hfil
  &\global\@eqcnt\tw@ $\displaystyle\tabskip\z@
  {##}$\hfil\tabskip\@centering&\llap{##}\tabskip\z@\cr}
\@addtoreset{equation}{section}
  \def\theequation{\thesection.\arabic{equation}}
\makeatother

\mathchardef\by="0202

\begin{document}

\title{Integrals of motion,
supersymmetric quantum mechanics and 
dynamical supersymmetry\footnote{\it 
Based on invited talk given at 
the International Seminar ``Supersymmetries and Quantum Symmetries"
dedicated to the memory of
Victor I. Ogievetsky (Dubna, July 22-26, 1997); to be published in
Proceedings.}} 
\author{Mikhail S. Plyushchay${}^{a,b}$\\
\smallskip\\
{\small ${}^a${\it Departamento de F\'{\i}sica, 
Universidad de Santiago de Chile,}}
{\small {\it Casilla 307, Santiago 2, Chile}}\\
{\small ${}^b${\it Institute for High Energy Physics,
Protvino, Moscow Region, Russia}}\\
{\small \it E-mail: mplyushc@lauca.usach.cl}}

\date{}

\maketitle

\begin{abstract}
The class of relativistic spin particle models reveals the
`quantization' of parameters already at the classical level.
The special parameter values emerge if one requires the
maximality of classical global continuous symmetries.  The same
requirement applied to a non-relativistic particle with odd
degrees of freedom gives rise to supersymmetric quantum
mechanics.  Coupling classical non-relativistic superparticle to
a `U(1) gauge field', one can arrive at the quantum dynamical
supersymmetry.  This consists in supersymmetry appearing at
special values of the coupling constant characterizing
interaction of a system of boson and fermion but disappearing in
a free case.  Possible relevance of this phenomenon to
high-temperature superconductivity is speculated.
\end{abstract}

\section{Introduction}
In ref. \cite{gps}
it was observed that the classical $(2+1)$-dimensional system 
given by Lagrangian~\cite{cpv}
\begin{equation}
L_f=\frac{1}{2e}(\dot{x}_\mu-iv\epsilon_{\mu\nu\lambda}
\xi^\nu \xi^\lambda)^2-\frac{e}{2}m^2
+2i\nu mv\theta_1\theta_2-\frac{i}{2}\xi_\mu\dot{\xi}{}^\mu
+\frac{i}{2}\theta_a\dot{\theta}_a
\label{l01}
\end{equation}
reveals a `classical quantization' of a real
parameter $\nu$.
The nature of this phenomenon is the following.
The model (\ref{l01}), described by even 
and odd  vectors $x_\mu$ and $\xi_\mu$, 
by scalar odd variables $\theta_a$, $a=1,2$,
and by even Lagrange multipliers $e$ and $v$,
admits the construction 
of oscillator-like odd variables $\xi^\pm$
linear in $\xi_\mu$, 
whose phases at $\vert\nu\vert=1$
evolve exactly as the phases of $\theta^\pm=\frac{1}{\sqrt{2}}
(\theta_1\pm i\theta_2)$.
As a consequence, two nilpotent integrals of motion
$\xi^+\theta^-$ and $\xi^-\theta^+$ can be 
constructed in addition to the integrals
$\theta^+\theta^-$ and $\xi^+\xi^-$.
So, at $\nu=\pm 1$ the system has 
a maximal classical global continuous symmetry.
The same values of $\nu$ are separated if one requires that the
system would have the maximal quantum global continuous symmetry.
Moreover, only at $\nu=\pm 1$ the discrete $P,T$-invariance 
of the classical system may be preserved at the quantum level.
As a result, the quantum model describes
the $P,T$-invariant system of massive fermions
realizing
irreducible representation of a nonstandard superextension of
the $(2+1)$-dimensional Poincar\'e group \cite{gps,gp}.

Instead of the pair of scalar Grassmann variables $\theta_a$, 
one can introduce
another Grassmann vector and construct the Lagrangian
similar to (\ref{l01}) \cite{np},
\begin{equation}
L_{CS}=\frac{1}{2e}(\dot{x}_\mu-\frac{i}{2}v\epsilon_{\mu\nu\lambda}
\xi^\nu_a \xi^\lambda_a)^2-\frac{e}{2}m^2
+2i\nu mv\xi_1^\mu\xi_{2\mu}-\frac{i}{2}\xi_{a\mu}\dot{\xi}_a{}^\mu.
\label{l02}
\end{equation}
It turns out that this system reveals
the same phenomenon of `classical quantization':
there are two special values of the parameter,
$\nu=\pm 2$, at which the system has a maximal global continuous symmetry.
In this case the corresponding additional classical local integrals of
motion are of the third order in Grassmann variables.
The same values of $\nu$ are separated at the quantum level 
from the requirement of the maximality of global continuous symmetry
and if to require preserving the classical invariance with respect to
the discrete $P$ and $T$ transformations.
Quantum mechanically, the system (\ref{l02})
describes the $P,T$-invariant system of Chern-Simons fields 
(topologically massive vector gauge fields \cite{cs}) \cite{np1,np2}.
Analogously to the fermion system
(\ref{l01}), the quantum states of the system (\ref{l02})
realize irreducible representation of a
nonstandard superextension of
the $(2+1)$-dimensional Poincar\'e group~\cite{np1,np2}.

Here we show that  
the requirement of the maximality
of classical global continuous symmetry
applied to a non-relativistic particle with odd degrees of
freedom gives rise naturally to supersymmetric quantum mechanics.
Then, coupling classical non-relativistic superparticle to  a `U(1)
gauge field', we arrive at the quantum dynamical supersymmetry.
The latter consists in supersymmetry appearing at special values of
the coupling constant characterizing interaction of a system of
boson and fermion but disappearing in a free case.  In
conclusion, we speculate on a  possible relevance of this
phenomenon to high-temperature superconductivity in the context
of a U${}_{\sigma_3}$(1) gauged version of the model (\ref{l01})
\cite{gps,gp}.

\section{Supersymmetric quantum mechanics}

Let us consider a nonrelativistic particle of unit mass
having two odd (Grassmann) degrees of 
freedom. Its  Lagrangian may be written 
in general form as
\begin{equation}
L=\frac{1}{2}\dot{x}{}^2-V(x)-R(x)N+
\frac{i}{2}\theta_a\dot{\theta}_a.
\label{lgen}
\end{equation}
Here $V(x)$ and $R(x)$ are two
arbitrary functions 
and $N=-i\theta_1\theta_2=\theta^+\theta^-$,
$\theta^\pm=\frac{1}{\sqrt{2}}(\theta_1\pm i\theta_2)$.
In the next section it will be shown that 
a possible inclusion of additional term
$\dot{x}{\cal A}(x)N$ does not change
the system (\ref{lgen}) classically but turns out
to be important quantum mechanically.
The nontrivial Poisson-Dirac brackets of the system
(\ref{lgen}) are
$\{x,p\}=1$, $\{\theta_a,\theta_b\}=-i\delta_{ab}$,
and the Hamiltonian 
$
H=\frac{1}{2}p^2+V(x)+R(x)N
$
generates the following equations of motion:
\[
\dot{x}=p,\quad 
\dot{p}=-V'(x)-R'(x)N,\quad 
\dot{\theta}{}^\pm=\pm iR(x)\theta^\pm.
\]
Thus, $N$ is an obvious integral of motion
additional to $H$.
If the evolution law $x=x(t)$ is known, then
the solution to equations of motion for odd variables 
is
\begin{equation}
\theta^\pm(t)=\theta^\pm(t_0)\exp\left[\pm i\int_{t_0}^t
R(x(\tau))d\tau\right],
\label{th+-}
\end{equation}
and we find that the odd quantities 
\begin{equation}
\Theta^\pm=\theta^\pm(t)\exp[\mp i\int_{t_0}^t
R(x(\tau))d\tau]
\label{nonloc}
\end{equation}
are {\it nonlocal in time} integrals of motion. 
In particular case $R=0$, odd variables have a trivial
dynamics, $\dot{\theta}{}^\pm=0$, and integrals 
(\ref{nonloc})  take the {\it local} form:
$\Theta^\pm=\theta^\pm$.
One may put the question:
is there any other nontrivial case
characterized by {\it local} odd integrals of motion instead of 
the nonlocal integrals (\ref{nonloc})?
It is clear that if the system
has even complex conjugate
quantities $A^\pm$, $(A^+)^*=A^-$,  whose 
evolution  looks up to the term proportional to $N$
like the evolution of odd variables in (\ref{th+-}),
then local odd integrals of motion could be constructed
in the form 
\begin{equation}
Q^\pm=A^\pm\theta^\mp.
\label{Qpm}
\end{equation}
Let us consider the oscillator-like variables
\begin{equation}
A^\pm=\frac{1}{\sqrt{2}}(\phi(x)\mp i p)
\label{apm}
\end{equation}
with some function $\phi(x)$.
We find that
\[
\dot{A}{}^\pm=\frac{1}{\sqrt{2}}
\left(\phi'p \pm i (V'+R'N)\right),
\]
and from the relation
\[
\dot{Q}{}^\pm=\pm i\left[(\phi'-R)Q^\pm+
\frac{1}{\sqrt{2}}(V'-\phi\phi')\theta^\mp\right]
\]
we conclude that $\dot{Q}{}^\pm=0$
when $\phi'=R$, $V'=\frac{1}{2}(\phi^2)'$.
Therefore, when the functions $R(x)$ and $V(x)$
are related as 
$
R(x)=W'(x),
$
$
V(x)=\frac{1}{2}W^2(x)+C,
$
where $W(x)=\phi(x)$ is an arbitrary function and $C$ is a
constant,  then odd quantities 
(\ref{Qpm}) are integrals of motion additional to 
$H$ and $N$.
Together with Hamiltonian $H$,
they form the classical analog of a central extension of 
$N=1$ supersymmetry algebra,
\[
\{Q^+,Q^-\}=-i(H-C),\quad 
\{H,Q^\pm\}=\{Q^+,Q^+\}=
\{Q^-,Q^-\}=0, 
\]
with constant $C$ playing a role of a central charge,
whereas the integral $N$ is classical analog
of the grading operator,
\[
\{N,Q^\pm\}=\pm iQ^\pm,\quad
\{N,H\}=0.
\]
Putting $C=0$, we arrive at the classical
analog of Witten's supersymmetric quantum mechanics \cite{wit}
given by the Lagrangian
\begin{equation}
L_{SUSY}=\frac{1}{2}\dot{x}{}^2-\frac{1}{2}W^2(x)+
i\theta_1\theta_2 W'(x)+\frac{i}{2}\theta_a\dot{\theta}_a.
\label{susy}
\end{equation}
We conclude that the system (\ref{susy})
is a special case of more general
classical system (\ref{lgen}),
which is characterized by the presence of two additional
{\it local} in time odd integrals of motion (\ref{Qpm})
being supersymmetry generators.

\section{Quantum dynamical supersymmetry}

Let us consider a system given by the Lagrangian
\begin{equation}
L_g=L_{SUSY}+\dot{x}{\cal A}(x)N.
\label{langl}
\end{equation}
Since $N$ is integral of motion of the system (\ref{susy}), 
one may write 
\begin{equation}
L_g=L_{SUSY}+\frac{d}{dt}\Delta,\quad
\Delta=N\int^{x(t)} {\cal A}(y)dy.
\label{total}
\end{equation}
The Hamiltonian for (\ref{langl}) is
\begin{equation}
H=\frac{1}{2}(p-{\cal A}(x)N)^2+\frac{1}{2}W^2(x)+N W'.
\label{hd}
\end{equation} 
In correspondence with lagrangian relation
(\ref{total}), the canonical transformation
\[
x\rightarrow \tilde{x}=x,\quad
p\rightarrow \tilde{p}=p-{\cal A}(x)N,
\quad
\theta^\pm\rightarrow 
\tilde{\theta}{}^\pm=\theta^\pm\exp(\pm i\varphi(x)),
\quad
\varphi(x)=\int_{}^x {\cal A}(y)dy
\]
reduces (\ref{hd}) to the Hamiltonian
of the `free' system  
(\ref{susy}) characterized by ${\cal A}(x)=0$.
Therefore, classical systems
(\ref{susy}) and (\ref{langl}) are equivalent
and the system (\ref{langl}) is supersymmetric 
for any ` U(1) gauge potential' ${\cal A}$.
However, due to ambiguities of quantization
procedure, the corresponding quantum
systems are not necessarily equivalent in generic case.
Let us show that using the 
quantization ambiguity, one can construct the
quantum system with supersymmetry appearing dynamically \cite{das}
at special values of a coupling constant
characterizing interaction of a system of
boson and fermion but disappearing 
in a free case.

We consider the simplest nontrivial 
system (\ref{hd}) given by ${\cal A}=g=const$ and $W(x)=\epsilon x$
with $\epsilon=1$:
\begin{equation}
H=\frac{1}{2}[(p-gN)^2+x^2]+\epsilon N.
\label{h0n}
\end{equation}
After realizing the canonical transformation
$x\rightarrow p$, $p\rightarrow -x$,
the Hamiltonian takes the form
\begin{equation}
H=\frac{1}{2}[p^2+(x+gN)^2]+ \epsilon N.
\label{en}
\end{equation}
This can be presented equivalently as 
\begin{equation}
H=\frac{1}{2}(p^2+x^2)+ \epsilon N +gxN,
\label{en1}
\end{equation}
and (\ref{en1}) can be understood as the Hamiltonian
of the system of
bosonic and fermionic oscillators
interacting through the coupling term $gxN$.
Quantum analog of (\ref{en1}) is
\begin{equation}
H=a^+a^-+\epsilon c^+c^- + gxc^+c^-,
\label{ds}
\end{equation}
where 
$a^\pm=\frac{1}{\sqrt{2}}(x\mp ip)$,
$[a^-,a^+]=1$, $[c^-,c^+]_{{}_+}=1$,
$c^{\pm2}=0$, $[a^\pm,c^\pm]=0$,
and passing from (\ref{en1}) to (\ref{ds})
we have chosen the normal ordering 
of bosonic and fermionic creation-annihilation operators.
Due to the quantum relation $N^2=N$, $N=c^+c^-$,
this quantum Hamiltonian can be written in the form
$H=b^+b^-+\tilde{\epsilon} c^+c^-$
with 
\[
b^\pm=a^\pm+\frac{g}{\sqrt{2}}c^+c^-,\quad
\tilde{\epsilon}=\epsilon-\frac{g^2}{2}.
\]
Performing in correspondence with classical picture
the unitary Bogoliubov transformation
$a^\pm\rightarrow b^\pm$,
$f^\pm=c^\pm\exp[\pm\frac{g}{\sqrt{2}}(a^- -a^+)]$, 
we present (\ref{ds}) equivalently as 
\begin{equation}
H=b^+b^-+\tilde{\epsilon} f^+f^-
\label{hff}
\end{equation}
with 
$[b^-,b^+]=[f^-,f^+]_{{}_+}=1$, $f^{\pm 2}=0$,
$[b^\pm,f^\pm]=0$.
The Hamiltonian
(\ref{hff}) is the exact quantum analog of the classical Hamiltonian
(\ref{en}), but 
with $\epsilon$ changed for
$\tilde{\epsilon}$. This difference happens since
we constructed (\ref{ds}) proceeding
from (\ref{en1}), and in contrast with  
classical relation $N^2=0$
quantum mechanically we have $N^2=N$.
The system (\ref{hff}) (and so, the system (\ref{ds})) is supersymmetric
at $\tilde{\epsilon}=1$, i.e. only in a free case ($g=0$). 
But we can interpret the system (\ref{ds}) in another
way. Suppose that the quantum system (\ref{ds}) 
is characterized by the parameter
$\epsilon=1+\frac{\alpha^2}{2}$ with
$\alpha\neq 0$, which means that bosonic and fermionic oscillators
have different frequencies.
Then in a free case ($g=0$)
this quantum system has no supersymmetry,
but turns into a supersymmetric system
at two special values of the coupling constant:
$g=\pm \alpha$.
If the coupling constant  
may be varied effectively as a function of some external parameter(s)
(temperature etc.), 
$g=g(\lambda)$, 
then the system will reveal supersymmetry
at special values of parameter(s)
defined by the relation 
$g^2(\lambda_s)=\alpha^2$.

\section{Concluding remarks}

For the models (\ref{l01}) and (\ref{l02}), the values of
corresponding parameters do not influence on the integrability
properties: the general solution to the equations of motion can
be constructed explicitly for arbitrary $\nu$s \cite{gps,np2}.  However,
there are special values of $\nu$s for which classical global
continuous symmetries of these systems happen to be maximal.
The situation is similar for the system (\ref{lgen}), but here
the requirement of maximality of classical global continuous
symmetry restricts and relates the form of two functions, $V(x)$
and $R(x)$.  If instead of the generic system (\ref{lgen}) we
consider its much more restricted case given by the Lagrangian
(\ref{susy}) with $W'(x)$ changed for $\nu W'(x)$, the
requirement of maximal classical symmetry
would lead us to the analogous `quantization condition':
$\nu^2=1$.  {}With this observation we arrive at the interesting
question whether the systems (\ref{l01}) and (\ref{l02}) may be
extended in such a way that the requirement of maximality of
classical global continuous symmetries would fix a functional
arbitrariness but not only the values of parameters.

The `U(1)-gauged' version of nonrelativistic
supersymmetric system is given by
Lagrangian (\ref{langl}). It is related classically to the
initial `free' system (\ref{susy}) by the canonical transformation, and
therefore also is supersymmetric.  On the other hand, the model
(\ref{l01}) may be
consistently coupled to a U${}_{\sigma_3}$(1) gauge field
(see refs. \cite{gps,gp}),
\begin{equation}
L_f\rightarrow
L_f^g=L_f+qN\left(\dot{x}_\mu{\cal A}^\mu+
\frac{e}{2}i\xi_\mu\xi_\nu{\cal F}^{\mu\nu}\right),
\label{fu1}
\end{equation}
where $q$ is a coupling constant,
${\cal A}_\mu={\cal A}_\mu(x)$ is a U(1) gauge field,
${\cal F}_{\mu\nu}=\partial_\mu{\cal A}_\nu
-\partial_\nu{\cal A}_\mu$,
and $N=\theta^+\theta^-$.
The structure of this $P,T$-invariant gauge interaction (see
ref. \cite{gps}) is similar to the structure of the `U(1) gauge
interaction' considered above for nonrelativistic supersymmetric
particle.  But now there is no trivial relationship of the
U(1)-gauged system (\ref{fu1}) to the initial 
free system (\ref{l01}) due to the presence in (\ref{fu1})
of the spin-field interaction term
proportional to $i\xi_\mu\xi_\nu{\cal F}^{\mu\nu}$. Thus,
it is not obvious what happens with classical global
continuous symmetries of relativistic spin particle system
(\ref{l01}) under coupling it to the U${}_{\sigma_3}$(1) gauge
field.  However, if one conjectures that the classical global
continuous symmetries of the system (\ref{l01}) survive in
the case of interaction (\ref{fu1}), what will happen with them at the
quantum level?  These questions on classical
and quantum symmetries of (\ref{fu1})
deserve further investigation. If 
the analog of the discussed phenomenon of quantum dynamical
supersymmetry also exists for the system (\ref{fu1}), 
this could
find important applications in the context of hypothetical
relevance of the $P,T$-invariant model (\ref{fu1}) 
to the problem of describing
high-temperature superconductivity
(see refs. \cite{gps,ht}).

\vskip1cm
{\bf Acknowledgement}
\vskip5mm

The work was supported in part by 
grant 1980619 from FONDECYT (Chile).


\begin{thebibliography}{**}

\bibitem{gps}
G. Grignani, M. Plyushchay and P. Sodano, 
{\it Nucl. Phys.} {\bf B464} (1996) 189,
hep-th/9511072.

\bibitem{cpv}
J. L. Cort\'es, M. S.  Plyushchay and L. Vel\'azquez,
{\it Phys. Lett.} {\bf B306} (1993) 34.

\bibitem{gp}
J. Gamboa and M. Plyushchay,
{\it Nucl. Phys.} {\bf B512} (1998) 485,
hep-th/9711170.

\bibitem{np}
Kh. S. Nirov and M. S. Plyushchay,
{\it Phys. Lett.} {\bf B405} (1997) 114,
hep-th/9707070.

\bibitem{np1}
Kh. S. Nirov and M. S. Plyushchay,
{\it J. High Energy Phys.} {\bf 02} (1998) 015,
hep-th/9712097.

\bibitem{np2}
Kh. S. Nirov and M. S. Plyushchay,
{\it Nucl. Phys.} {\bf B512} (1998) 295, hep-th/9803221.

\bibitem{cs}
R. Jackiw and S. Templeton, {\it Phys. Rev.} 
{\bf D23} (1981) 2291;\\
J. Schonfeld, {\it Nucl. Phys.} {\bf B185} (1981) 157;\\
W. Siegel, {\it Nucl. Phys.} {\bf B156} (1979) 135;\\
S. Deser, R. Jackiw and S. Templeton,
{\it Phys. Rev. Lett.} {\bf 48} (1982) 975.

\bibitem{wit}
E. Witten, {\it Nucl. Phys.}
{\bf B188} (1981) 513; {\bf B202} (1982) 253.

\bibitem{das}
A. Das and M. Hott, 
hep-th/9504059.

\bibitem{ht}
A. Kovner and B. Rosenstein, {\it Phys. Rev.}
{\bf B42} (1990) 4748;\\
G.W. Semenoff and N. Weiss,
{\it Phys. Lett.} {\bf B250} (1990) 117;\\
N. Dorey and N.E. Mavromatos,
{\it Nucl. Phys.} {\bf B386} (1992) 614.

\end{thebibliography}
\end{document}